\documentclass[fleqn,twoside]{article}
\usepackage{espcrc2}
\usepackage{amssymb}
\usepackage{graphicx}
\usepackage[figuresright]{rotating}
\newcommand{\half}{\frac{1}{2}}
\newcommand{\none}{\multicolumn{1}{c}{--}}

\title{Spectrum of SU(2) SUSY Yang-Mills Theory with a light gluino}

\author{Roland Peetz\address[MS]{Institut f\"ur Theoretische Physik,
        Universit\"at M\"unster, Wilhelm-Klemm-Str.9, D-48149
        M\"unster, Germany}, Federico Farchioni\addressmark[MS], 
        Claus Gebert\address{Deutsches Elektronen Synchrotron DESY, 
        Notkestr. 85, D-22607 Hamburg, Germany}, Gernot
        M\"unster\addressmark[MS]}

\begin{document}

\begin{abstract}
We report on new results for the low lying spectrum of N=1
SUSY Yang-Mills Theory with SU(2) as the gauge group. Simulating on larger
lattices at $\kappa = 0.194$ and $\beta = 2.3$, we slowly approach the
supersymmetric limit at $m_{gluino} = 0$.
\end{abstract}

\maketitle

\section{Introduction}
Progress has been made in further determining the low lying spectrum
of N=1 SU(2) SUSY Yang-Mills Theory (SYM). The motivation for doing large
scale simulations has been presented at previous conferences and 
in various publications, for a review on the subject see \cite{Montvay:2001aj}.

This work is related to a previous project of the DESY-M\"unster-Roma
collaboration to simulate SYM in the vicinity of
the supersymmetric point \cite{Montvay:2001aj} using the 
dynamical-fermion two step multibosonic algorithm (TSMB). 
This involves a light gluino mass.
Previous calculations left open the questions 
a) how close to the SUSY limit was the model actually simulated
and b) the multiplet structure of the spectrum. 
The first question was answered by last year's SUSY Ward-Identity results 
\cite{Farchioni:2001wx} showing that the gluino was heavier than expected
from previous estimates. The second question is still under investigation
and up to now not conclusively answered.
Here we attempt an analysis of the spectrum closer to the SUSY limit, i.e.
with a lighter gluino, where the SUSY pattern of masses should become more 
apparent.

\section{N=1 SYM on the lattice}

A basic assumption about the non-perturbative behavior of N=1 SYM is
confinement, as in QCD. Following results from effective action
analyses \cite{Farrar1997fn}, we expect to see 
two chiral multiplets at the bottom of the SYM mass spectrum. 
Their content is a spin-$\half$ fermion (the gluino-glueball) 
and two bosonic states with opposite parity: a scalar and
pseudoscalar gluino-gluino bound state (gluinoball) $a-f_0$ and $a-\eta'$
({\em a} denoting 
the adjoint representation of the gluinos) and the scalar and
pseudoscalar glueball with $J^{PC} = 0^{-+}$ and $0^{++}$.

On the lattice, Poincar\'e invariance is broken and therefore SUSY. 
We use Wilson fermions where SUSY is also explicitely broken 
by the lack of chiral symmetry. Finally a soft breaking is 
caused by the gluino mass. 
Because of all this SUSY breaking the SYM spectrum gets distorted 
on the lattice in an essentially uncontrolled way.

We compute the SYM mass spectrum on the lattice from first 
principles using the familiar techniques known in QCD.
The correlation functions of the gluinoballs are similar to
those of QCD flavor singlets 
\begin{eqnarray}
  C_{\tilde g \tilde g}(\Delta t)=\sum_{\vec{x}} \left< 
    {\rm Tr_{sc}}[\Gamma \Delta_{xx}] {\rm Tr_{sc}}[\Gamma \Delta_{yy}]
    \right.\nonumber \\ \left.
      \quad \quad -2{\rm Tr_{sc}} [\Gamma \Delta_{xy}\Gamma\Delta_{yx}]
    \right>.  \nonumber
\end{eqnarray}
where $y$ is a fixed source, the trace is over spin and color and $\Gamma \in
(1,\gamma_5) $.

The gluino-glueball is associated in the continuum with the interpolating 
operator $\phi=\sigma_{\mu \nu}{\rm Tr_c}[F_{\mu \nu}\lambda]$
where $\lambda(x)$ is the gluino field; on the lattice the correlation is 
\begin{eqnarray}
C_{\tilde g g}^{\alpha \beta}(\Delta
t)=-\frac{1}{4}\sum_{\vec{x}}\sum_{i,j,k,l,\alpha^\prime,\beta^\prime}\sigma_{ij}^{\alpha \alpha'}  
{\rm Tr_c}[U_{ij}(x)\sigma^{a}] \Delta_{xa;yb}^{\alpha'\beta'}
\nonumber \\ \times {\rm Tr_c}[U_{kl}(y)
\sigma^{b}]\sigma_{kl}^{\beta'\beta}.\quad \nonumber
\end{eqnarray}
where the trace here is only over color.
The glueball operators are the equivalents to those of QCD.

In order to be able to make meaningful statements of the ermergence of SUSY 
in the continuum through extrapolations, we want the gluino to be 
as light as possible. As in QCD, the natural barrier 
is algorithmic performance due to critical slowing down. 

\section{Numerics}

Following Curci and Veneziano, we employ 
\[ S_{eff}[U] = \beta \sum_P \left( 1- \half  {\rm Re Tr } U_P) \right)- 
\half \log\det Q \]
($Q$ is the usual fermionic matrix) from which we see that the gluino 
effectively has flavor number $N_f=\half$.  Expectation values of operators read
\[ \left< \mathcal{O} \right> = Z^{-1}\int {\cal D[}U] {\cal PF}[M] {\cal
O}(U) e^{-S_{gauge}[U]} \] 
where ${\cal PF}[M]$ is the Pfaffian of the antisymmetric fermion matrix
$M=CQ$. 

The most suitable algorithm for this model is the two-step
multiboson (TSMB) algorithm \cite{Montvay:1995ea};
for details on its application to SYM see \cite{Campos:1999du}. 
It relies on representing the fermion determinant in the form 
\[ |\det(Q)|^{N_f} \simeq \frac{1}{\det P_{n_1}^{(1)}(\tilde{Q}^2)
P_{n_2}^{(2)}(\tilde{Q}^2)}\quad. \]
The polynomial approximations satisfy
\begin{eqnarray}
        P_{n_1}^{(1)}(\tilde{Q}^2) &\simeq& x^{-N_f/2} \nonumber \\
        \lim_{n_2 \rightarrow \infty} P_{n_1}^{(1)}(x)P_{n_2}^{(2)}(x)
        & = & x^{-N_f/2}, \quad x\in[\epsilon,\lambda]\nonumber 
\end{eqnarray}
where the eigenvalues of $\tilde{Q}^2=Q^{\dagger}Q$ on a typical gauge
configuration are required to be in the interval $[\epsilon,\lambda]$.
$P_{n_1}^{(1)}$ gives a crude estimate of the
fermionic measure and is used in the bosonic representation of the
determinant. $P_{n_2}^{(2)}$  is a correction factor that is accounted for by 
a global accept-reject step. The algorithm is made exact by
a third polynomial $P_{n_3}^{(3)}$  through reweighting the gauge 
configurations in the expectation values.

For the results presented we use the following samples, all at $\beta=2.3$:
\begin{center}
  \begin{tabular}{lclll}\hline
    \multicolumn{1}{c}{$\kappa$} & 
    \multicolumn{1}{c}{$L\times T$}  &
    \multicolumn{1}{c}{$\epsilon$} & 
    \multicolumn{1}{c}{$\lambda$} & 
    Stat  
    \\ \hline
   0.1925 &  12$\times$24  & $0.0003$      & 3.7 & 4204 \\ 
   0.194  &  12$\times$24  & $0.0001$      & 4.5 & 2034 \\ 
   0.1955 &  12$\times$24  & $0.0000125$  & 5.0 & 5324 \\ 
   0.194  &  16$\times$32  & $0.0002$ & 4.0 & 664 \\
    \hline 
  \end{tabular}
\end{center}
where we also indicate the $\epsilon$ and $\lambda$ used in the simulations.
The configuration on the $12^3\times24$ lattice were produced in 
\cite{Campos:1999du} and \cite{Farchioni:2001wx}.
In order to check finite size effects we started a new production
run on a $16^3\times 32$ lattice; the configurations on the last line
of the above table are a snapshot of it.

From the SUSY Ward-identities study \cite{Farchioni:2001wx} it follows 
$\kappa_c(\beta=2.3) \approx 0.197$, 
where $\kappa_c(\beta)$ is the hopping parameter where the gluino is massless.

\begin{table*}
\caption{Summary of our spectrum results. New results are in bold face.}
\begin{center}
\renewcommand{\tabcolsep}{1.14pc}
\begin{tabular}{lllllll}\hline
 \multicolumn{1}{c}{$\kappa$} & 
 \multicolumn{1}{c}{$L \times T$}&
 \multicolumn{1}{c}{$am_{0^{++}}$} & 
 \multicolumn{1}{c}{$am_{0^{-+}}$}  & 
 \multicolumn{1}{c}{$am_{a-\eta'}$} & 
 \multicolumn{1}{c}{$am_{f_0}$} &
 \multicolumn{1}{c}{$am_{g \tilde{g}}$}
 \\\hline
 0.1925 & $12 \times 24$ & 0.53(10)       &{\bf 0.80(18)}  & 0.48(5) &
 1.00(13)& {\bf 0.883(16)} \\   
 0.194  & $12 \times 24$ & {\bf 0.40(11)} &{\bf 1.10(28)}  & \none   & \none
 & {\bf 0.816(18)} \\   
 0.1955 & $12 \times 24$ & \none          &   \none        & \none   &  \none
 & {\bf 0.751(21)} \\   
 0.194  & $16 \times 32$ & \none          &   \none        &{\bf 0.49(6)} &
 \none& \none\\  
 \hline
\end{tabular}    
\label{tab:summary}
\end{center} 
\end{table*}

\section{Results}

The new results presented here essentially refer to the two 
samples of configurations on the $12^3\times24$ lattice, 
$\kappa=0.194$, $0.1955$ produced in \cite{Farchioni:2001wx}. 
These correspond to a lighter gluino compared to the previous extensive
analysis of the spectrum accomplished in \cite{Campos:1999du} where the
largest $\kappa$ value was $0.1925$. We also present
preliminary results for the bigger $16^3\times32$ lattice at $\kappa=0.194$.
The available results are summarized in table \ref{tab:summary}.

\vspace{.1cm}

\begin{figure}[htb]
\begin{center}
\vspace{3mm}
\includegraphics[angle=0,width=0.46\textwidth]{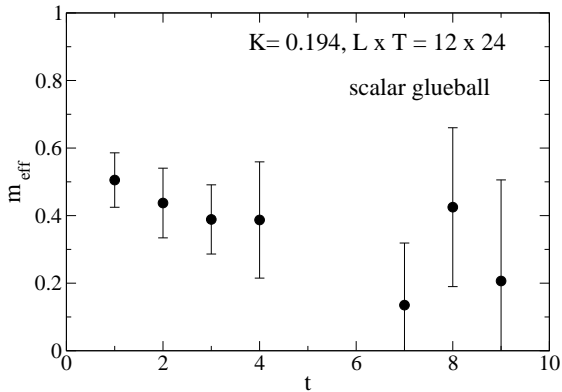}
\vspace{-9mm}
\caption{\small $m_{eff}$ of the $0^{++}$-glueball}
\label{zpglueball}
\end{center}
\end{figure}

\noindent
{\bf Glueballs.}
We used APE smearing with $N_{APE} = 36,40$ and $\epsilon_{APE}=0.285$ for the
glueball operators. We then averaged the effective masses over the emerging
plateau. The data for the $0^{-+}$-glueball is generally very noisy so that
averages can only be taken at time-slice sparations 1 and 2. The case of the
$0^{++}$-glueball is more favorable (see fig.~\ref{zpglueball}). The
statistics for the larger lattice is still too low to determine a mass.

\vspace{.1cm}

\begin{figure}[htb]
\begin{center}
\vspace{3mm}
\includegraphics[angle=0,width=0.46\textwidth]{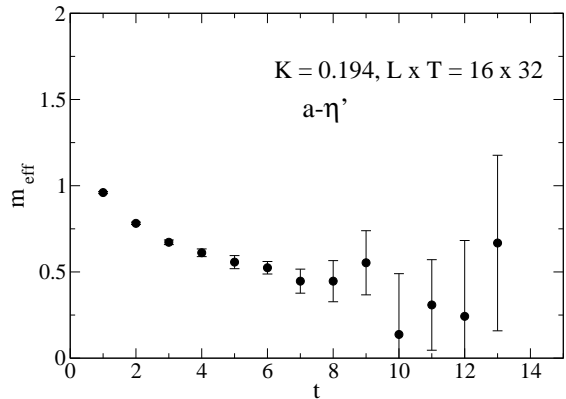}
\vspace{-9mm}
\caption{\small $m_{eff}$ of the $a-\eta'$-gluinoball}
\label{aetap}
\end{center}
\end{figure}

\noindent
{\bf Gluinoballs.}
The connected part of the correlation function was evaluated by choosing a
random source, the disconnected by using the volume-source technique, both
without smearing. The signal for the $a-\eta'$ is reasonably good (see 
fig.~\ref{aetap}) whereas for the $a-f_0$ no mass could be extracted yet.

\vspace{.1cm}

\noindent
{\bf Gluino-glueballs.}
The gluino-glue correlation $C^{\alpha \beta}(\Delta t)$ is evaluated by
calculating the propagator $\Delta^{\alpha \beta}_{xy}$ for a random source
$y$.  APE as well as Jacobi smearing for the gluino have been used with
parameters ($N_{APE} =9$, $\epsilon_{APE} = 0.5$) and ($N_{Jac} = 18$,
$\epsilon_{Jac}= 0.2$). 

\vspace{.1cm}

\noindent
{\bf The pseudo chiral limit.}
Within the OZI-approximation the connected term in the $a-\eta'$ correlator is
expected to give rise to a massless mode in the limit $m_{\tilde{g}}
\rightarrow 0$, the ``adjoint pion" $a-\pi$ \cite{Veneziano:1982ah}.
Extrapolating values for $m_{a-\pi}$ at various $\kappa$'s, 
we find $\kappa_c \simeq 0.1970(4)$, which is consistent with the
determination in  
\cite{Farchioni:2001wx}.

\section{Conclusions and outlook}

We still have work to do to fill out the blank spots in
table \ref{tab:summary}. 
We plan to use better numerical technology such as variational smearing
for glueballs, and stochastic estimators and spectral decomposition
of the fermion propagator for the disconnected parts of the gluino-balls
correlations. It would also be interesting to investigate the
effects of the volume on the spectrum. For this better statistics are needed.
\newline 

The computations were carried out on the Cray T3E at NIC, J\"ulich and the Sun
Fire SMP-Cluster at RWTH Aachen, Germany.


\begin{thebibliography}{9}

\bibitem{Montvay:2001aj}
  I.~Montvay, Int.\ J.\ Mod.\ Phys.\ A {\bf 17} (2002) 2377.
\bibitem{Farchioni:2001wx}
  F.~Farchioni {\it et al.}  [DESY-Munster-Roma Collaboration],
  Eur.\ Phys.\ J.\ C {\bf 23} (2002) 719.
\bibitem{Farrar1997fn}
  G.~R.~Farrar, G.~Gabadadze and M.~Schwetz,
  Phys.\ Rev.\ D {\bf 58} (1998) 015009.
\bibitem{Montvay:1995ea}
  I.~Montvay,
  Nucl.\ Phys.\ B {\bf 466} (1996) 259.
\bibitem{Campos:1999du}
  I.~Campos {\it et al.}  [DESY-Munster Collaboration],
  Eur.\ Phys.\ J.\ C {\bf 11} (1999) 507.
\bibitem{Veneziano:1982ah}
  G.~Veneziano and S.~Yankielowicz,
  Phys.\ Lett.\ B {\bf 113} (1982) 231.
\end{thebibliography}
\end{document}